\UseRawInputEncoding
\documentclass{optica-article}

\journal{opticajournal} 

\articletype{Research Article}

\usepackage{lineno}

\begin{document}

\title{Wafer-scale waveguide sidewall roughness scattering loss characterization by image processing}

\author{Mohit Khurana\authormark{1,2,*}, Sahar Delfan\authormark{1,2}, Zhenhuan Yi\authormark{1,2,\textdagger}}
\address{\authormark{1}Department of Physics and Astronomy, Texas A\&M University, College Station, TX 77843, USA\\
\authormark{2}Institute for Quantum Science and Engineering, Texas A\&M University, College Station, TX  77843, USA\\}
\email{\authormark{*}mohitkhurana@tamu.edu, \authormark{\textdagger}yzh@tamu.edu} 


\begin{abstract*}Photonic integrated circuits (PICs) are vital for developing affordable, high-performance optoelectronic devices that can be manufactured at an industrial scale, driving innovation and efficiency in various applications. Optical loss of modes in thin film waveguides and devices is a critical measure of their performance. Thin films growth, lithography, masking, and etching processes are imperfect processes that introduce significant sidewall and top-surface roughness and cause dominating optical losses in waveguides and photonic structures. These roughness as perturbations couple light from guided to far-field radiation modes, leading to scattering losses that can be estimated from theoretical models. Typically, with UV-based lithography sidewall roughness is found to be significantly larger than wafer-top surface roughness. Atomic force microscopy (AFM) imaging measurement gives 3D and high-resolution roughness profile but the measurement is inconvenient, costly, and unscalable for large-scale PICs and at wafer-scale. Here, we evaluate the sidewall roughness profile based on 2D high-resolution scanning electron microscope imaging. We characterized the loss on two homemade nitride and oxide films on 3-inch silicon wafers with 12 waveguide devices on each and co-related the scattering loss estimated from a 2D image-based sidewall profile and theoretical Payne model. The lowest loss of guided fundamental transverse electric (TE$_{0}$) is found at 0.075 dB/cm at 633 nm across 24 devices, which is a record at visible wavelength. Our work shows a 100\%  success in edge detection in image processing to estimate autocorrelation function and optical mode loss.  These demonstrations offer valuable insights into waveguide sidewall roughness and comparison of experimental and 2D SEM image-processing-based loss estimations.

\end{abstract*}

\section{Introduction}
Photonic integrated circuits (PICs) are vital in advancing various technologies, including high-speed optical communication, optical interconnects, PIC-based optical clocks, neural networks-based optical computing, sensing, light sources, and quantum circuits \cite{Shekhar2024, Newman2019, Chauhan2021, Margalit2021}. Manufacturing these devices on a large scale requires precise and scalable fabrication techniques and processes for high-performance yield. The critical components of these procedures are lithography, patterning, and etching of material to develop the photonic components on the chip. The primary techniques are photolithography and electron-beam lithography (EBL), with photolithography being the predominant choice in the industry. Although it has resolution limitations, it is generally preferred for all applications except prototypes because of its speed, cost-effectiveness, and scalability. In the large-scale manufacture of PICs, photolithography enables the simultaneous patterning of entire wafers, rapidly producing hundreds or thousands of PIC components within minutes. In atomic and molecular research applications, resolutions within the 50-100 nm range are generally sufficient for PIC architectures and components in visible and near-infrared wavelengths, making deep UV photolithography more practical\cite{Cui2024}. In contrast, EBL offers higher resolution and can achieve features as small as 5 nm, but it is significantly expensive and slower, taking hours or even days for a single wafer. EBL allows design flexibility without requiring new masks; its high operational costs make it better suited for prototyping and small-scale production.
\\

\noindent 
The resolution of a lithographic process is determined by a simplified version of the Rayleigh criterion, $R = \frac{k_1 \cdot \lambda}{NA}$, where, \( R \) is the smallest resolvable feature size (resolution), \( \lambda \) is the wavelength of the light source, \( NA \) is the numerical aperture of the imaging system, \( k_1 \) is a process constant dependent on the resist and the patterning technique (typically around 0.4 to 0.9 in practical photolithography processes), the resolution improves with shorter wavelengths (\(\lambda\)) and higher numerical apertures (NA) \cite{Mack2007}. In photolithography, ongoing research and advancements aim to reduce the costs of small wavelength laser sources and hardware components, enhancing device yield performance. The 365 nm (i-line) wavelength works well for older semiconductor devices, while the 248 nm and 193 nm wavelengths accommodate increasingly smaller features, achieving down to 65 nm for advanced photolithography \cite{He2023}. The state-of-the-art 13.5 nm (EUV) technology allows for features patterning as tiny as 10 nm, underscoring the industry's commitment to developing more intricate circuits. In visible-telecom wavelengths, PICs necessitate minimum feature sizes for components like waveguides, microring resonators, on-chip modulators, photonic crystal waveguides, and cavities, roughly around 50 nm. 
\\

\noindent 
Thin film growth, lithography patterning, masking, and etching processes are imperfect processes that introduce significant sidewall and top-surface roughness and cause dominating optical losses in waveguides and photonic structures. These roughnesses act as perturbations that couple light from guided to far-field radiation modes, leading to unwanted scattering losses. Waveguide loss is evaluated as the total loss quantity, $\alpha$ includes all the effects such as radiative loss, surface and sidewall roughness loss, absorption in material, material defects, mechanical defects, etc. Sidewall roughness is significantly larger than wafer-top surface roughness due to high-quality thin film, low-quality resist edges, and top-down etching process. Scattering losses from sidewall ($\alpha_{sidewall}$) and surface ($\alpha_{surface}$) roughness are typically the most dominant factors among all for a straight rectangular waveguide. Evaluating loss due to this roughness using Maxwell's solvers like FDTD is impractical and is measured directly from experimental observations. Experimental methods include high-resolution roughness measurements using atomic force microscopy and optical characterization of waveguide loss. Experiments measure average loss, i.e., $\alpha = \alpha_{avg.} = \frac{\sum_{i}\alpha _{i} L _{i}}{\sum_{i}L_i}$, where $\alpha_{i}$ is optical loss of $L_i$ waveguide length unless advanced imaging and optical characterization are performed such as confocal microscopy - which is quite a challenging task because the loss of photons is minimal for typical used optical powers, slow speed measurement and require additional experimental verifications with high accuracy. Loss measurement is a significant factor in characterizing PIC components. As such, the Q-factor of an on-chip integrated cavity is estimated from total loss quantity, $\alpha$ \cite{Roberts2022}. For the microring resonator, the total loss is evaluated as $\alpha_{ring} = 2\pi n_{g}/(Q\lambda) = \alpha_{mat.} + \alpha_{scat.} + \alpha_{TIR}$; in the case of non-symmetric edges of waveguides, $\alpha_{scat.}=\alpha_{side1}+\alpha_{side2}$. Similarly, for photonic crystal cavity the optical loss, $\alpha=\alpha_{TIR} +\alpha_{mat.} +\alpha_{scat.}$ and the Q factor is estimated by, $1/Q_{tot.} = 1/Q_{TIR} + 1/Q_{mat.} + 1/Q_{scat.}$ \cite{Englund2006}. Note that, $\alpha_{scat.} > \alpha_{TIR}>\alpha_{mat.}$ therefore, $1/Q_{tot.} = 1/Q_{TIR} + 1/Q_{scat.}$.
\\

\noindent For a PIC with thousands of components and a large number of PICs in industry-scale manufacturing, characterization of loss quantity with direct optical measurements is not feasible. As the next photonics revolution awaits, a quick and efficient measurement technique of the optical loss quantity is needed. Electron microscope images can extract features sub-10 nm size characteristics of devices and have been used as feedback in the FDTD simulator to estimate actual photonic device performance \cite{Englund2006}. However, FDTD is a computationally time-consuming and intensive method to estimate scattering loss and evaluate device performance. A combined methodology of quick measurement of the roughness of waveguide edge based on high-resolution imaging and loss estimation using a theoretical model would be advantageous in characterizing micro and macro parameters, optical losses, and the performance of PIC devices. 2D (electron microscopy) and 3D (atomic force microscopy, focused ion beam tomography, confocal microscopy) imaging methods are crucial in analyzing surface features and sidewall structures. SEM is considerably high-resolution, cheaper, faster, and more convenient; however, it overlooks the chiseling effect (variation over the depth of waveguide edge roughness), though this is minimal due to the top-down etching process. The sem resolution limit, $R_{\text{SEM}} = \frac{C_s \cdot \lambda}{\alpha}$, where $C_s$ is the spherical aberration coefficient of the lens,  $\lambda$ is the wavelength of the electrons and $\alpha$ is the semi-angle of the electron beam convergence. At typical SEM operating voltages (10-100 kV), the electron wavelength \( \lambda \) is on the order of picometers, making it possible to achieve high resolution \cite{Tanaka2008}. However, aberrations such as spherical and chromatic aberrations degrade the resolving power of the system. Non-conductive samples require coating with a thin conductive layer (e.g., gold or carbon) to prevent charging effects, which can obscure fine details.
\\

\noindent
In this research investigation, we perform image processing on SEM images of waveguide edges to characterize the roughness, use a theoretical model to empirically estimate the waveguide loss, and compare it with the experimentally observed loss at a wafer-scale level.


\section{Theoretical loss and waveguide edge roughness}
The analytical expression of radiation loss coefficient for scattering by surface roughness in a symmetric single-mode waveguide that is based on coupled-mode theory (by Marcuse in 1969) was reported by Payne et al.  \cite{Payne1994, Marcuse1969}, 
\begin{equation}
\alpha_{cm^{-1}} = \frac{L_{dB/cm}}{4.343} = \varphi^2(d) (n_1^2 - n_2^2)^2 \frac{k_0^3}{4 \pi n_1} \int_0^\pi \tilde{R}(\beta - n_2 k_0 \cos \theta) \, d\theta 
\end{equation}
and scattering loss based on three-dimensional models using coupled-mode theory (2006) and volume current method (2005) have been reported by Poulton et al. and  Barwicz et al., respectively \cite{Poulton2006, Barwicz2005}. These models have accurately predicted losses in low and high-contrast refractive index core waveguides and cladding. In Eq. 1, if one wall is rough, then $\alpha$ must be multiplied by 1/2, $\varphi(d)$ is the modal field evaluated at the waveguide surface and is normalized so that, $\int_{-\infty}^{\infty} \varphi^2(y) \, dy = 1$; $n_1$ and $n_2$ are the core and cladding refractive indices, respectively; $k_0$ is the free-space wavenumber and $\beta$ is the modal propagation constant; the width of the waveguide is $2d$. For the bent waveguide, the loss, $\alpha= \alpha_{side1}$ + $\alpha_{side2}$, where the mode field at both sides is to be estimated numerically from the EME or FDTD method. The surface roughness of the waveguide walls is described by the spectral density function $\tilde{R}(\Omega)$, which is related to the autocorrelation function (ACF) $R(u)$ of the surface roughness through the Fourier transform, $\tilde{R}(\Omega) = \int_{-\infty}^{\infty} R(u) \exp(i \Omega u) \, du$ \cite{Payne1994, Marcuse1969}. The roughness described by R(u) is a function of f(z) - which is a one-dimensional distribution with zero mean, $R(u)=\left\langle\ f(z)f(z+u) \right\rangle$, where the brackets represent ensemble average, in case of the straight waveguide, f(z) is a variation of edges around the ideal straight waveguide \cite{Poulton2006, Barwicz2005}. In the case of a bent waveguide, the edge curve of the waveguide can be best fitted to a polynomial function. Now, the models mostly used for ACF that closely relate to the roughness produced by photolithography, hard-mask, and top-down etching processes in the fab are: $R(u) = \sigma^2 \exp \left( -\frac{|u|}{L_c} \right)$, $R(u) = \sigma^2 \exp \left( -\frac{u^2}{L_c^2} \right)$ and $R(u) = \sigma_{1}^2 \exp \left( -\frac{|u|}{L_{c}} \right) + \sigma_{2}^2 cos(2\pi u/D)$. The surface roughness is characterized by a correlation length $L_c$, and mean square deviation $\sigma^2$ from a flat surface, where $\sigma^2 = R(0)$. In waveguide roughness, ACF is useful in understanding spatial dependencies in surface or edge variation. The ACF measures how a roughness profile correlates with itself at different location lags. Other models include, $R(u) = A_0 e^{-\frac{u}{D_1}} + A_1 e^{-\frac{u}{D_2}}$, $R(u)= A_0 e^{-\gamma u} \cos(\omega u)$, $R(u) = A_0 e^{-\left( \frac{u}{L_0} \right)^\beta}$ (\(0 < \beta \leq 1\)), $R(u) = A_0 - \alpha \log(u)$, etc. The analytical integral expression, $S = \int_0^\pi \tilde{R}(\beta - n_2 k_0 \cos \theta) \, d\theta$ for all these different ACFs may not be available, but numerically, it can be estimated.
\\

\noindent
Image processing of SEM images to extract ACF involves a crucial step of precise edge detection. Various edge detection algorithms can identify image boundaries based on sudden shifts in pixel intensity, including Canny, Sobel, Prewitt, Laplacian of Gaussian, and Difference of Gaussians (DoG), etc. We discuss these algorithms here briefly. Canny's algorithm is a multi-stage algorithm that enhances accuracy through noise reduction, gradient computation, and hysteresis thresholding; however, it requires careful parameter tuning 
\cite{Canny1986}. Sobel utilizes two 3x3 convolution kernels to approximate gradients in horizontal and vertical directions, offering computational efficiency and some noise reduction but struggles in low-contrast environments. The Prewitt operator is similar but lacks weight for the central pixel, making it less accurate and more noise-sensitive. The Laplacian of Gaussian (LoG) combines Gaussian smoothing with the Laplace operator to detect edges through zero-crossings but can be compromised by noise if not correctly adjusted. Each technique presents a unique balance of accuracy, noise sensitivity, and computational complexity, catering to diverse computer vision and image analysis applications. The Canny algorithm is the most accurate for general use and offers excellent noise reduction and edge localization. The Canny's algorithm consists of applying a Gaussian filter to smooth the image, $ G(x, y) = \frac{1}{2\pi \sigma^2} e^{ -\frac{x^2 + y^2}{2\sigma^2} }$ and convolving the image \( I(x, y) \) with \( G(x, y) \) reduces high-frequency noise, computation of the intensity gradients using operators like Sobel filters, non-maximum suppression by thin the edges to one-pixel width by suppressing non-maximum gradient magnitudes, double thresholding by applying two thresholds to classify pixels as firm edges, weak edges, or non-edges and edge tracking by hysteresis to finalize edge detection by connecting weak edges that are linked to solid edges \cite{edge_detection}. To accurately capture image details, the sampling frequency must satisfy the Nyquist criterion: $f_s \geq 2 f_{\text{max}}$, where \( f_s \) is the sampling frequency, and \( f_{\text{max}} \) is the highest frequency present in the image. Higher resolution (higher \( f_s \)) allows for accurate representation of high-frequency details, which is crucial for precise edge detection. The effectiveness of this algorithm is influenced by image resolution and pixel count. Higher resolution typically enhances the accuracy of imaging systems; however, it also contributes to increased computational complexity ($T_{\text{Canny}} \propto N = N_x \times N_y$). If an image has dimensions \( N_x \) pixels in width and \( N_y \) pixels in height, the total pixel count \( N \) is: $N = N_x \times N_y$. The physical size of each pixel depends on the imaging sensor or sampling process. The spatial resolution \( R \) (in pixels per unit length) can be expressed as: $R_x = \frac{N_x}{L_x}, \quad R_y = \frac{N_y}{L_y}$, where \( L_x \) and \( L_y \) are the physical dimensions of the image along the x and y axes, respectively. The pixel size \( \Delta x \) and \( \Delta y \) (physical distance represented by each pixel) are: $\Delta x = \frac{1}{R_x} = \frac{L_x}{N_x}, \quad \Delta y = \frac{1}{R_y} = \frac{L_y}{N_y}$.  The gradient of the image intensity \( I(x, y) \) is a key component in edge detection, defined as: $\nabla I = \left( \frac{\partial I}{\partial x}, \frac{\partial I}{\partial y} \right)$. In discrete images, gradients are approximated using finite differences: $\frac{\partial I}{\partial x} \approx \frac{I(i+1, j) - I(i-1, j)}{2\Delta x}, \quad \frac{\partial I}{\partial y} \approx \frac{I(i, j+1) - I(i, j-1)}{2\Delta y}$. Smaller \( \Delta x \) and \( \Delta y \) (higher resolution) lead to more accurate gradient estimations, improving edge detection accuracy \cite{Solomon2010}.

\section{Results and Discussion}
\begin{figure}[ht!]
\centering\includegraphics[width=15.5cm]{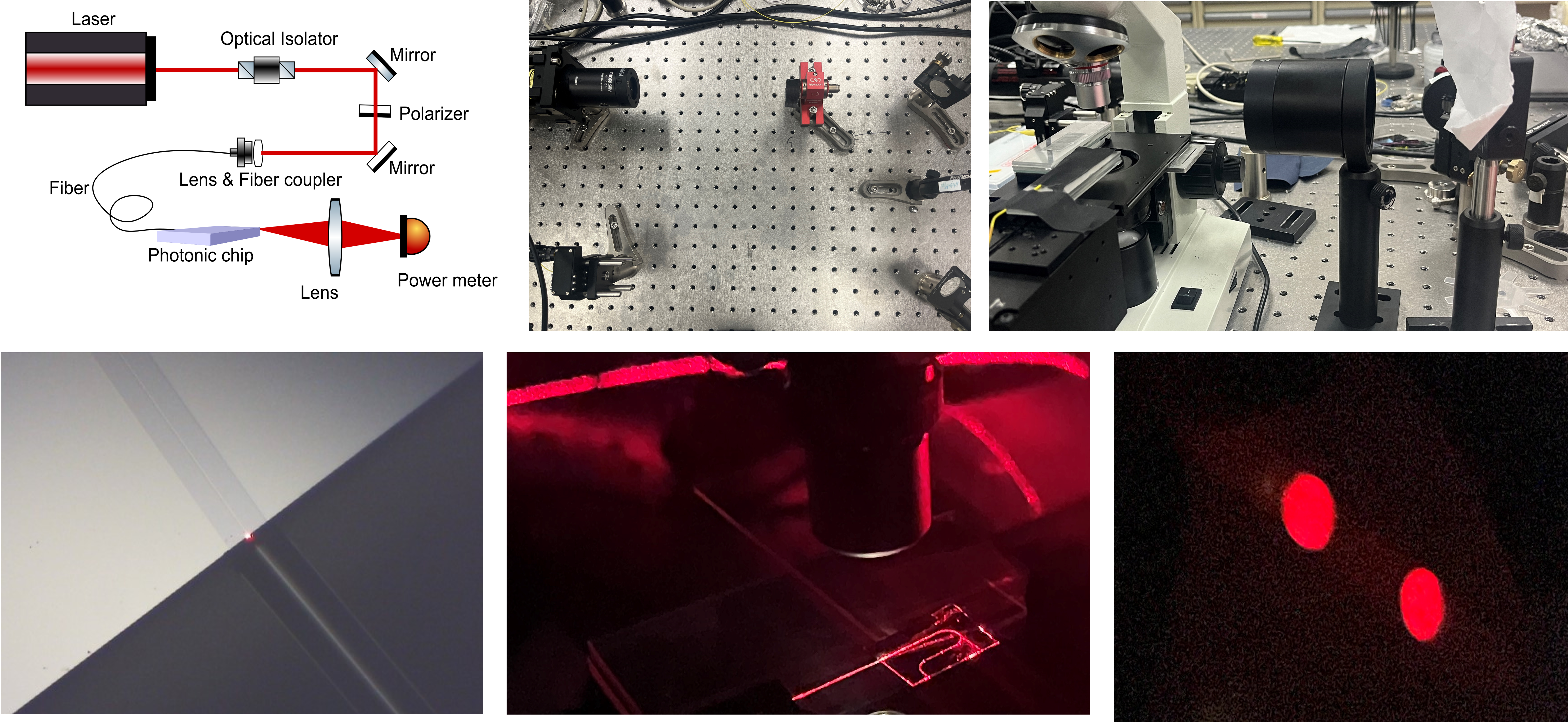}
\caption{Experimental setup for photonic waveguide loss characterization. Laser is launched through an optical isolator and polarized and then coupled to fiber that is butt-coupled to a photonic chip waveguide; output diverging beams pass through a lens of focal length 8 cm, and each beam is then captured in a photodetector (power meter). a) Schematic diagram of the experimental setup, b) top-view image of optical setup, c) Stage-chip and lens system, d) Captured image from a camera to collect reflected light from the objective-stage system shows SM fiber butt-coupled to a waveguide chip, e) Side-view image in a dark room with light coupled to waveguide chip with an objective-stage system and f) two output beams observed on white paper screen. In a typical experiment, $\sim 5\,\mu W$ is coupled in to waveguide TE$_{0}$ mode.} 
\label{setup_experimental}
\end{figure}

To investigate and demonstrate the estimation of optical loss at a wafer-level scale using a theoretical model combined with high-resolution sem image processing, we fabricated two silicon nitride waveguide wafers, each containing 12 waveguide devices designed for characterizing optical loss in waveguides. Each device includes a single optical waveguide that splits into two waveguide branches, with one waveguide branch intentionally designed to have a longer length. This variation in path lengths and precise output power measurements facilitate the quantification of optical losses. The measurement process involves coupling a continuous-wave laser into the input waveguide using a high-precision alignment setup, ensuring optimal coupling efficiency as shown in Fig. \ref{setup_experimental}. As the light propagates through the two branches, losses due to scattering, absorption, and fabrication imperfections accumulate along the path length. At the output, each branch's light intensity is measured using a photodetector - power meter. The power difference between the two outputs directly correlates with the optical loss per unit length, as the longer waveguide accumulates proportionally greater loss. To ensure accuracy, the two waveguide branches' length difference is identical for each device, and multiple measurements are conducted to account for variations in input coupling and environmental factors.

\begin{figure}[ht!]
\centering\includegraphics[width=15.5cm]{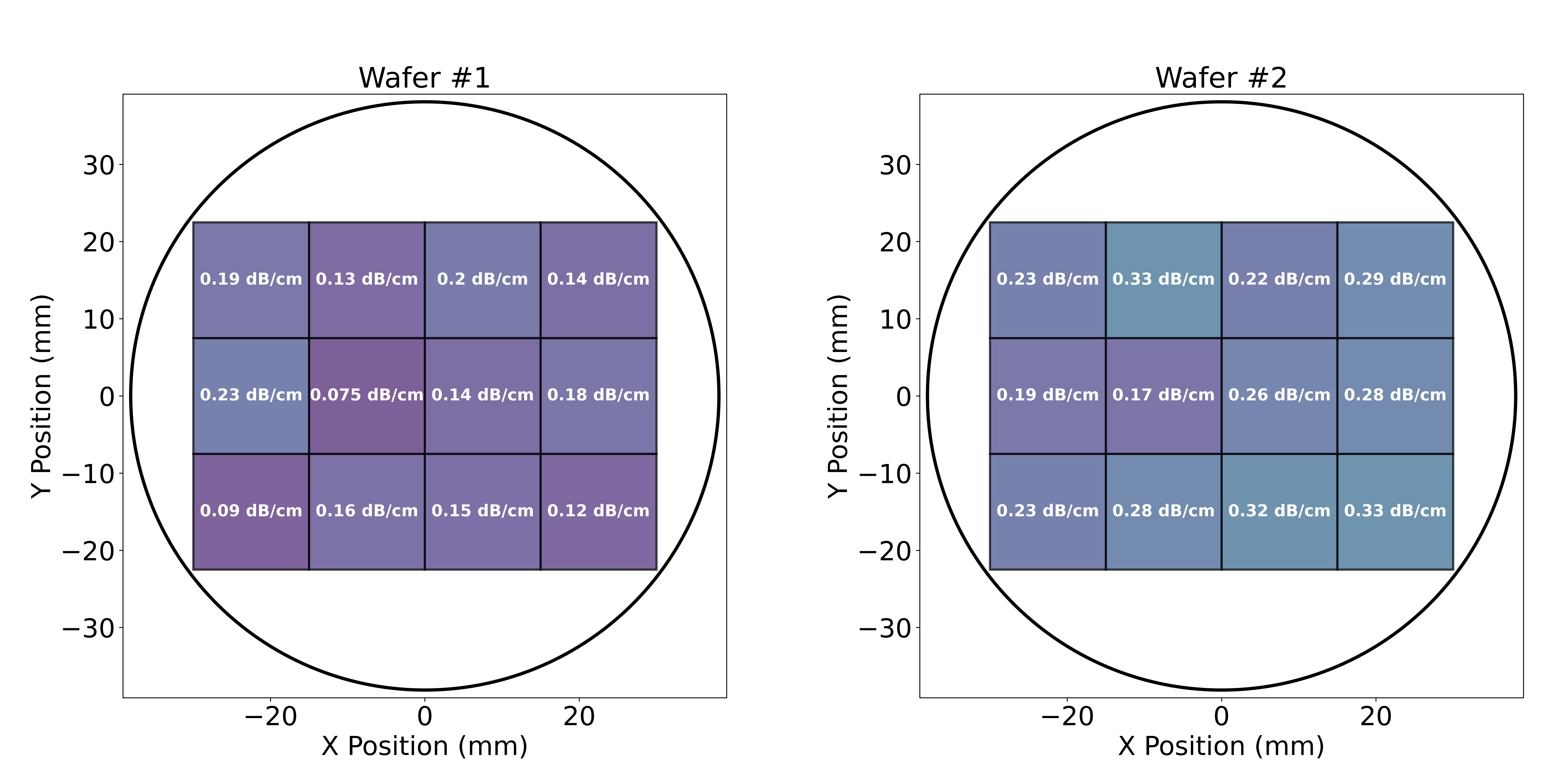}

\caption{Waveguide loss wafer profile, the experimental observed waveguide optical loss of each device on two wafers (total devices = 24). For all measurements, the error in values is less than 10\%.}
\label{wafer_profile}
\end{figure}

\noindent In intermediate steps of image processing and experimental characterization, first, the waveguide is patterned with the lithography and etching process followed by deposition of a thin Gold (Au) nanoparticles layer for acquiring sem images. A thin Au layer was deposited on the wafer to achieve high-contrast images. After obtaining the images, the Au layer was etched away using an Au etchant, followed by a cleaning procedure. After that, a silica layer (upper cladding) with a thickness of 1 $\mu$m is deposited on the wafer. Each wafer was cut into 12 individual dies using a diamond cutter, each representing one device. Each die was placed under an optical setup as illustrated in Fig. \ref{setup_experimental}. The waveguide architecture included a Y-splitter that divided the waveguide into two outputs. These outputs were collected on a white screen or power meter for loss characterization of the devices. The optical loss for each device was estimated and presented in Fig. \ref{wafer_profile}. The observed optical loss ranged from 0.18 to 0.36 dB/cm for wafer 1, while for wafer 2, it ranged from 0.39 to 0.73 dB/cm across the 12 devices. We believe the loss discrepancy between the two wafers can be attributed to variations in the fabrication processes conducted on different days. The supplementary information presents details of fabrication, experimental procedure, image processing, edge detection algorithms, and optical loss estimation.
\\

\begin{figure}[ht!]
\centering\includegraphics[width=15.5cm]{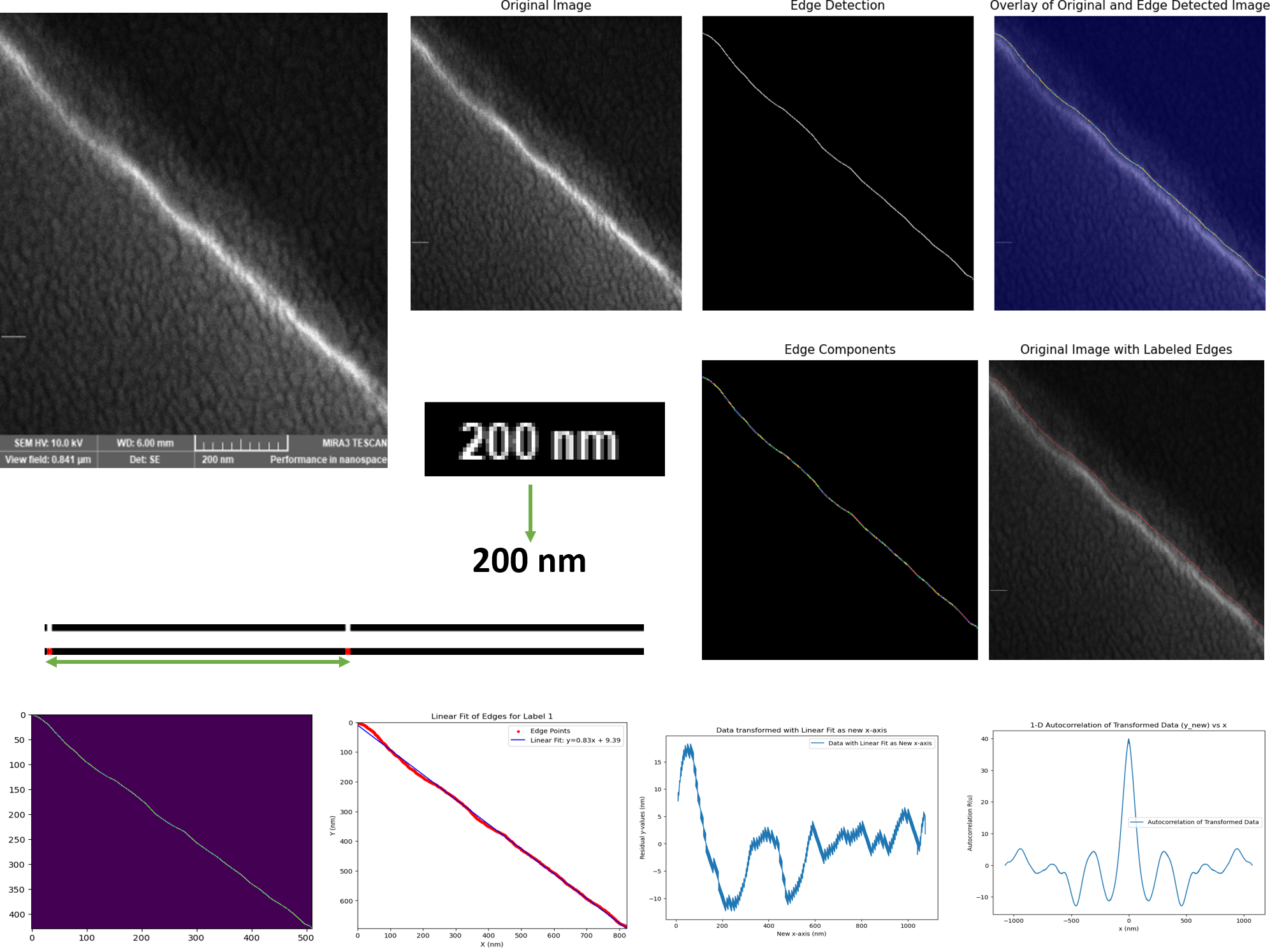}
\caption{Electron microscope image processing. a) The original image acquired from the sem tool. b) Trimmed image. c) Edge detection utilizes a sigmoid correction and an optimized parameter process for the edge detection algorithm. d) Overlay of both the original and edge-detected images for comparative visualization. e) Quantitative extraction of the scalebar and corresponding scale length measured in pixels from the image. f) Plot illustrating the detected edges corresponding to pixel numbers. g) Edge data normalized to a metric scale, accompanied by a linear fit of the data, represented in blue. h) Transformation of data aligned along the fit line. i) Autocorrelation plot analyzing the data presented in (h), highlighting the periodicity and relationships within the transformed dataset.}
\label{edge_detection_process}
\end{figure}

\noindent To estimate the edge roughness by edge detection processing on sem images, we collected 90 sem images of the waveguide edge sections from each of the fabricated wafers. An illustration of image processing performed on SEM images is shown in Fig. \ref{edge_detection_process} to extract the waveguide edges' ACF and roughness parameters. First, the pixel scale information is extracted from the original image. The pixel scale in nanometers is converted by fitting the endpoints of the scalebar to its numerical value within the image. Following this, the image underwent sigmoid correction, and an edge detection algorithm was used to identify the waveguide edges. These detected edges were then fitted to an ideal expected straight line, and the resulting ACF was plotted.
\\

\begin{figure}[ht!]
\centering\includegraphics[width=15.5cm]{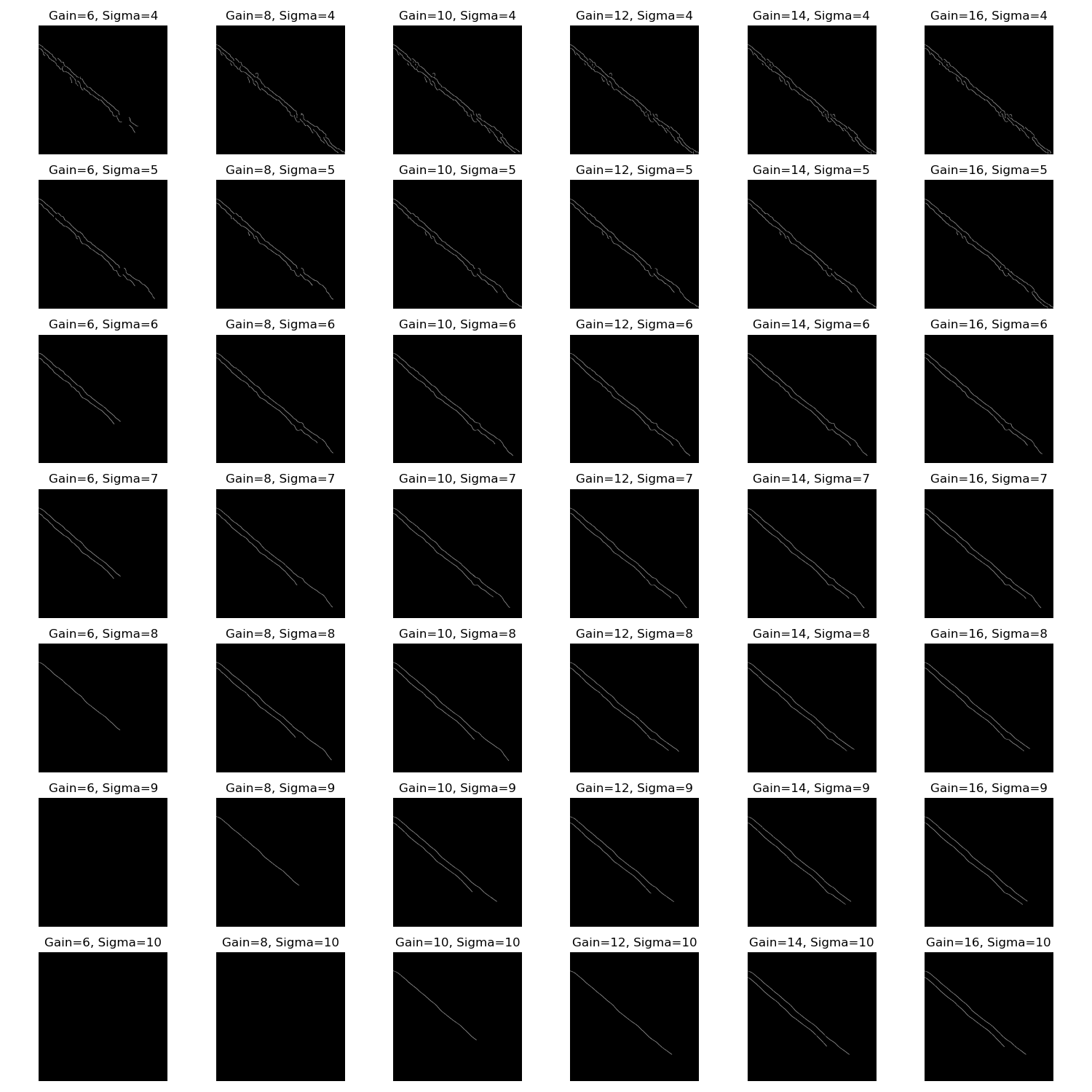}
\caption{Edge detection optimization process involves image filtering with a sigmoid correction gain parameter and an edge detection sigma parameter. This method applies various values of gain and sigma to enhance edge detection, which are critical for achieving high-quality, continuous edges that accurately represent the waveguide boundary.}
\label{all_edge_detections}
\end{figure}

\begin{figure}[ht!]
\centering\includegraphics[width=15.5cm]{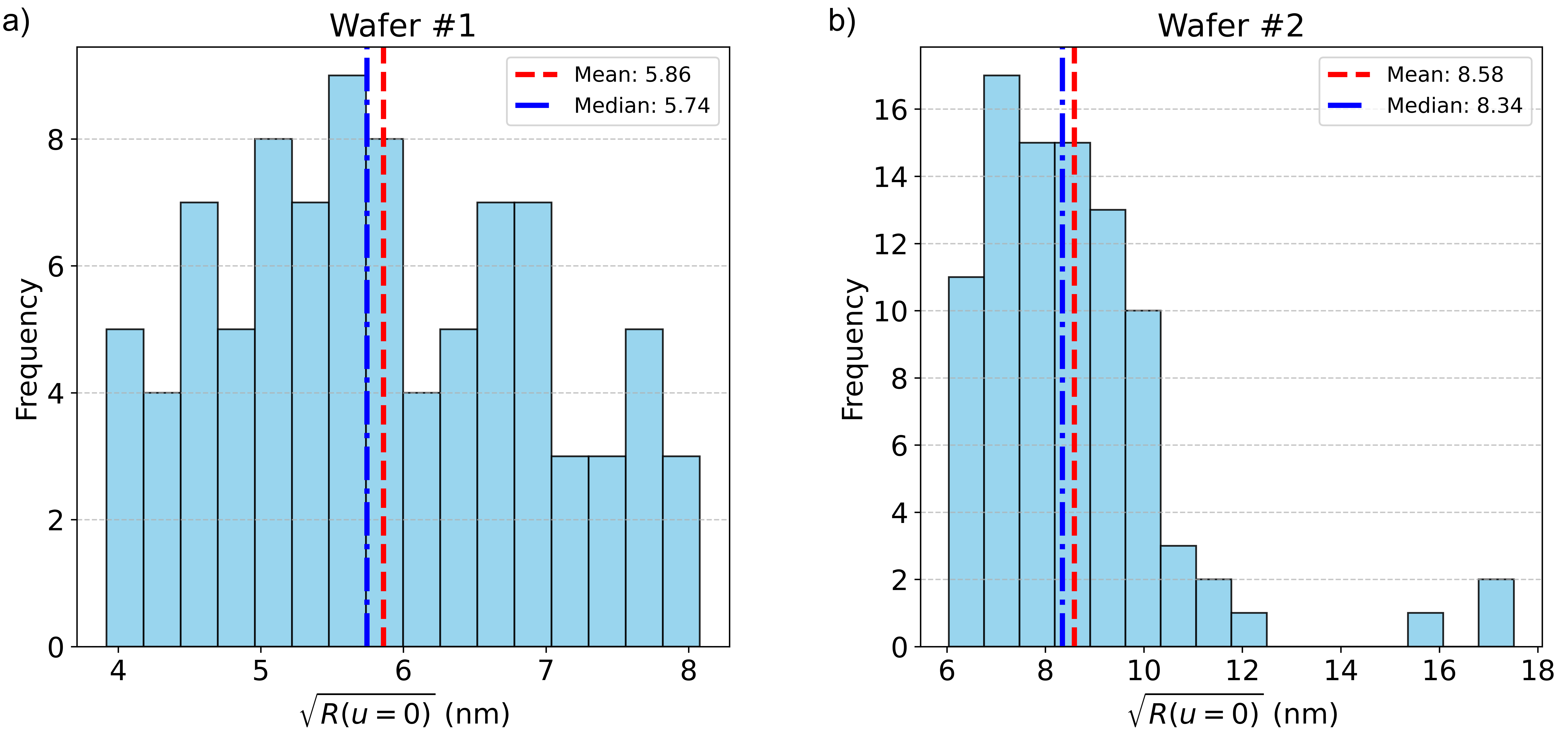}
\caption{Statistical analysis of the waveguide edge roughness of two wafers based on ACF extracted from methods discussed before, the x-axis is roughness (in nm), and the y-axis is the total number of edges or sem images (one edge per sem image) with a bin.}
\label{array_roughness}
\end{figure}

\noindent Fig. \ref{all_edge_detections} shows the result of edge detection algorithm parameters variation on a SEM image. Sigmoid correction gain and canny edge detection sigma parameters vary, affecting the final edge detection. This indicates the need to optimize these two parameters to find single-edge detection at the edge. We performed a particle-swarm optimization process and maximized the edge length to achieve 100 percent success in detecting an edge for a given waveguide SEM image. We evaluated ACF for each 90 images randomly acquired from a wafer, in total 180 images for two wafers. Fig. \ref{array_roughness} shows the histograms of the computed $\sqrt{R(u=0)}$ values derived from 90 SEM images of the respective wafers - (a) and (b) corresponding to two wafers. The results indicate that the observed differences in losses are linked to increased sidewall roughness of the waveguides, as evidenced by the variations in the computed $\sqrt{R(u=0)}$ values between the two wafers. The mean $\sqrt{R(u=0)}$ for the first wafer was 5.86 nm, while for the second wafer, it measured 8.58 nm. Notably, the first wafer exhibited a spread of up to 2 nm distribution around its mean value. In contrast, the second wafer displayed a distribution of approximately 3-4 nm, with three values falling within the 15-17 nm range. The extracted values from fitted data characterize the waveguide sidewall roughness, \(L_c\) \( \sim 100 \, \text{nm}\), a mean \(\sigma\) of \(5.86 \, \text{nm}\) and \(8.58 \, \text{nm}\)  (wafer 1 and 2 respectively), $\lambda$ \( \sim 633 \, \text{nm}\), an effective refractive index (\(n_{\text{eff}}\)) of \(1.5\), and refractive indices \(n_1 = 2.0\) and \(n_2 = 1.46\), along with a phase-related parameter \(\phi_d = 0.3\). Using these inputs, the free-space wavenumber is calculated as $k_0 = \frac{2\pi}{\lambda}$ and the propagation constant is given by $\beta = n_{\text{eff}} \cdot k_0$. See supplementary material for the theoretical expression of S. Based on the calculation, the loss is 0.08 dB/cm and 0.16 dB/cm for wafers 1 and 2, respectively.
\\

\clearpage

\begin{figure}[ht!]
\centering\includegraphics[width=15.5cm]{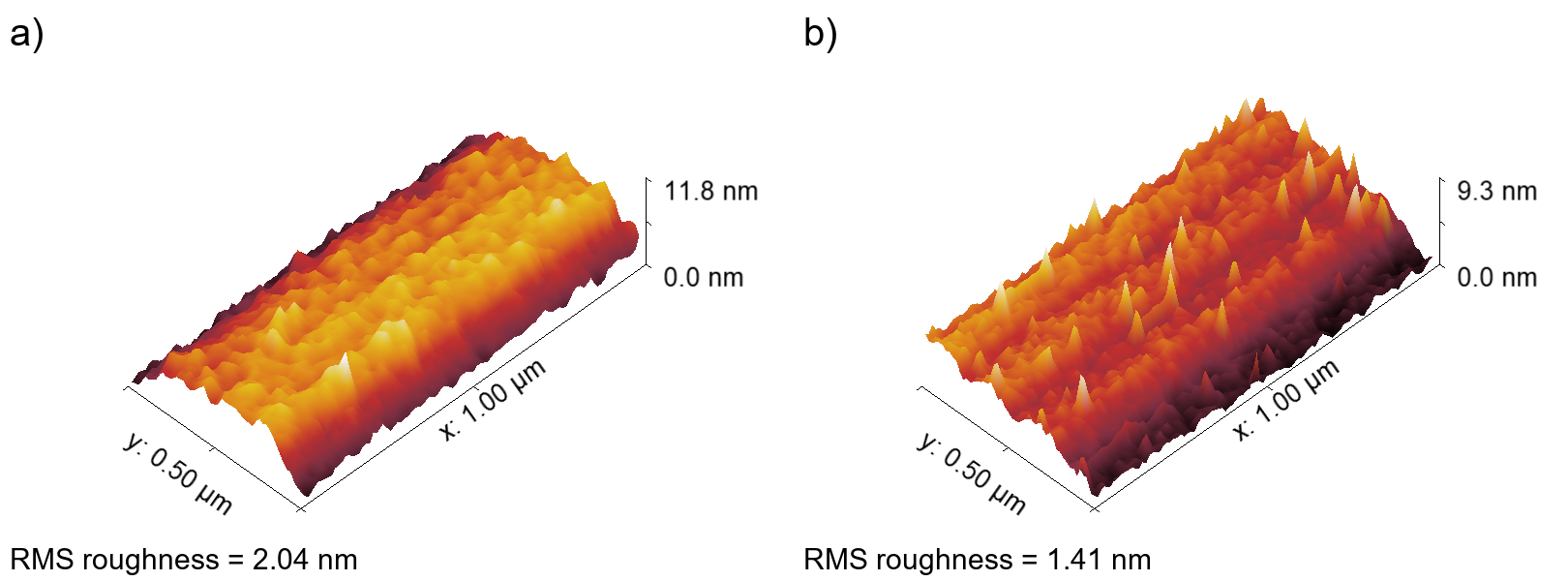}
\caption{AFM surface profiles of silicon nitride thin films. (a) The surface profile of a homemade silicon nitride thin film with a thickness of 40 nm shows an RMS roughness of 2.04 nm over a 1.00 $\mu$m $\times$ 0.50 $\mu$m scan area. (b) The surface profile of a commercially bought silicon nitride thin film with a thickness of 200 nm shows an RMS roughness of 1.41 nm over a 1.00 $\mu$m $\times$ 0.50 $\mu$m scan area. The color scale indicates height variation, demonstrating differences in surface quality between the homemade and commercial nitride films.}
\label{afm_array}
\end{figure}

\noindent We further report the comparison of surface roughness between commercially bought stoichiometric LPCVD nitride film (Rogue Valley Microdevices) and our homemade fabricated LPCVD nitride films (see Fig. \ref{afm_array}) based on surface AFM measurements. The AFM tool used in this measurement has vertical (Z) and horizontal (XY) axes resolution $\sim$ 1 nm and $\sim$ 10 nm, respectively. The commercially available silicon nitride thin film of thickness 200 nm demonstrates a lower root mean square (RMS) roughness of 1.41 nm compared to the home-manufactured silicon nitride thin film of thickness 40 nm, exhibiting an RMS roughness of 2.04 nm as shown in Fig. \ref{afm_array}. Nevertheless, the surface roughness of the home-manufactured film remains within an acceptable range. It is pertinent to consider that both surface roughness values are relatively low when juxtaposed with the typical edge roughness observed in side-wall roughness of waveguides as observed, which exceeds three to four times these values. This observation suggests that, in the context of optical losses, the surface roughness of these films is unlikely to be the predominant factor when compared to the inherent roughness at the edges of waveguides. Nevertheless, in both cases, the surface roughness is several times less than the edge roughness; therefore, sidewall roughness losses dominate the total optical waveguide loss demonstrated in this work.

\section{Discussion and Conclusion}
Modeling sidewall roughness and scattering losses are vital in estimating the waveguide loss, and there is a need to reduce the complexity in an optical characterization procedure to calculate the optical losses of photonic waveguide-based devices\cite{KuanPeiYap2009, Roberts2022} so that the process can be scaled to a large number of devices in a convenient and efficient way. In this study, we examined SEM images taken at various locations of straight waveguide edge sections and derived the values of Lc and sigma through ACF fitting data. The Lc and sigma values differ within the fabricated devices across 3-inch PICs and two distinct wafers. 
\\

\noindent This work can potentially be applied to characterize various photonic devices, such as the Q-factor of on-chip microring cavities, the Q-factor of photonic crystals, and the losses of waveguide crossing optical modes. Since the intrinsic Q-factor of cavities is approximated based on the total and where sidewall roughness scattering prevails, sidewall roughness scattering loss equals the total loss, as discussed in the introduction.
\\

\noindent Various theories for modeling roughness-induced scattering loss in rectangular waveguides exhibit significant variations in loss estimations\cite{Payne1994, Poulton2006, Barwicz2005}. However, the definite understanding is that roughness (ACF and roughness, sigma factor) contributes to scattering-induced radiation loss or light loss from a waveguide. Without loss of generality, we can characterize photonic waveguide-based devices and waveguide losses by evaluating S factor, where $S = \int_0^\pi \tilde{R}(\beta - n_2 k_0 \cos \theta) \, d\theta$. For instance, in waveguide crossing, comparison of S values of different waveguide crossings, a relative loss estimate can be made if not absolute value. Additionally, from the perspective view for future work, machine learning can be effectively utilized, a predictive framework can be established by training a model on experimental or simulated data correlating roughness parameters -such as amplitude, correlation length, and standard deviation - with waveguide loss. This approach would address the computational challenges associated with traditional numerical simulations in complex PIC architectures. The trained model can generalize the roughness-loss relationship by incorporating features like waveguide dimensions, material properties, and roughness metrics, facilitating rapid predictions across diverse designs. This methodology could potentially be a valuable tool for optimizing PIC performance and minimizing losses, circumventing the need for costly and time-intensive full-scale simulations or experimental assessments.
\\

\noindent In this work, we demonstrated wafer-scale silicon nitride waveguide optical loss characterization with sem image processing and compared it with experiments. We tested 12 devices on each of two wafers, and 0.075 dB/cm was the lowest loss observed at 633 nm. We demonstrated low-loss waveguides on homemade silicon nitride and oxide films at 633 nm. This experimental result adds a new record in previously reported low-loss nitride films for photonic waveguides and devices. We demonstrated the application of edge detection algorithms to characterize the edge roughness and evaluated optical loss employing a theoretical model. The statistics of waveguide edge roughness over 90 sem images for each wafer are demonstrated. The minimum resolution of roughness estimation achieved in our work is $\sim$ 4 nm, which is not limited by a fundamental principle. The work presented here would be valuable in potentially characterizing the photonic devices on large-scale PIC efficiently and quickly, which is a far better option than AFM imaging based roughness estimation. 

\section{Author Contributions}
M.K. and S.D. designed and performed the optical experiment, and S.D. performed the fabrication and acquired sem and AFM images of the samples. M.K. conceived the idea, performed the analytical studies, image processing, and data analysis, and wrote the manuscript. Z.Y. contributed to discussions.


\section{Funding}
S.D. is supported by Herman F. Heep and Minnie Belle Heep Texas A\&M University Endowed Fund held/administered by the Texas A\&M Foundation. We want to thank the Robert A. Welch Foundation (grants A-1261 and A-1547), the DARPA PhENOM program, the Air Force Office of Scientific Research (Award No. FA9550-20-10366), and the National Science Foundation (Grant No. PHY-2013771). This material is also based upon work supported by the U.S. Department of Energy, Office of Science, Office of Biological and Environmental Research under Award Number DE-SC-0023103, DE-AC36-08GO28308. 

\section{Acknowledgment}
We thank Prof. Marlan O. Scully for the insightful discussions during this study. The photonic chip was fabricated at Texas A\&M University's Aggiefab facility. We would like to thank Ming-Hsun Chou for assisting Sahar with the AFM tool.

\section{Additional information}
Supplement material is included as a separate file for additional information.

\section{Disclosures}
The authors declare that they have no known competing financial interests or personal relationships that could have appeared to influence the work reported in this paper.

\section{Data Availability}
Data underlying the results presented in this paper are not publicly available at this time but can be obtained from the corresponding author upon reasonable request.


\bibliography{sample}

\end{document}